\documentclass[multphys,vecphys]{svmult}

\usepackage{makeidx}         % allows index generation
\usepackage{graphicx}        % standard LaTeX graphics tool
\usepackage{multicol}        % used for the two-column index
\usepackage[bottom]{footmisc}% places footnotes at page bottom
\makeindex             % used for the subject index
\def\unit#1{\;{\rm #1}}
\def\Teff{\hbox{$T_{\rm eff}$}}
\def\logg{log\hspace*{1mm}$g$}
\def\SW{${\rm \Sigma W}$}
\def\VVhb{$(V_{\rm HB}- V)$}

\begin{document}

\title*{The Age-Metallicity Relation of the SMC}
\author{Andrea Kayser\inst{1}\and
Eva K. Grebel\inst{1}\and
Daniel R. Harbeck\inst{2}\and
Andrew A. Cole\inst{3}\and
Andreas Koch\inst{1}\and
John S. Gallagher\inst{2}\and
Gary S. Da Costa\inst{4}
}
\authorrunning{Kayser et al.}
\institute{Astronomical Institut, U. Basel, Venusstr. 7, 4102 Binningen, Switzerland
\texttt{akayser@astro.unibas.ch, grebel@astro.unibas.ch}
\and Dept. Astronomy, U. Wisconsin, 475 N. Carter St. Madison, WI 53706, USA
\texttt{}
\and Astronomy Dept., U. Minnesota, 116 Church St. SE, MN 55455, USA
\texttt{}
\and ANU, Mt Stromlo Observatory, Cotter Rd, Weston ACT 2611, Australia
\texttt{}}
\maketitle

\section{Introduction}
\label{sec:1}
The Small Magellanic Cloud (SMC), as one of our nearest galactic neighbours, provides a very important laboratory for the study of galaxy evolution.
Its proximity enables us to resolve stellar populations well below the oldest main sequence turn-offs (MSTO) and thus allows reliable age dating.
Moreover the star cluster (SC) system of the SMC shows no sign of any substantial age gap, as found, e. g., in the LMC. In fact it is the only dwarf galaxy in the Local Group known to have formed and preserved populous SCs continuously over the past $\sim$12 Gyr. This provides a unique, closely spaced set of single-age, single-metallicity tracers.

In a former study, spectroscopic metallicities for 6 populous SMC clusters with ages from 3 to 12 Gyr were obtained by \cite{DC/H:98}.
They found that the chemical evolution of the SMC was generally consistent with a simple "closed-box" model and a smoothly varying star-formation rate. However, this conclusion has been challenged by studies based on additional photometric data \cite{M:98} suggesting bursty evolution, or even a bimodal age distribution \cite{R:00}. 

All existing age-metallicity relation studies suffer from uncertainties due to either small number statistics or photometry based abundances.
Moreover, deep high resolution HST imaging data are available for only a subset of SCs in the SMC.
Thus all present age determinations are based on different studies of varying quality.

\section{Data and Analysis}
\subsection{Data}
\label{sec:2.2}
\begin{figure}[t!]
\centering
\begin{tabular}{c c}
\hskip-0.1cm\includegraphics[height=5.1cm]{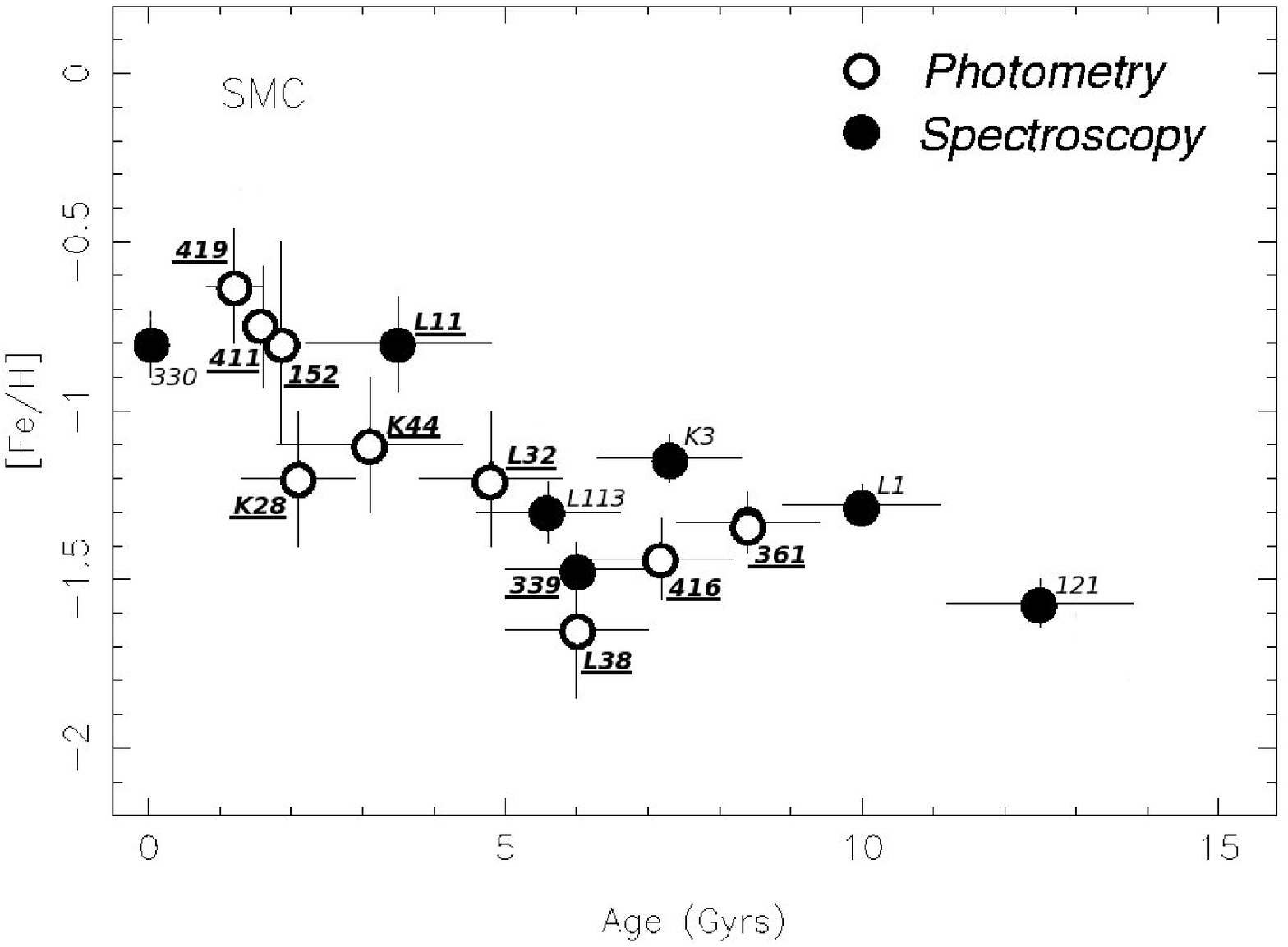} &
\hskip-0.1cm\includegraphics[height=5.1cm]{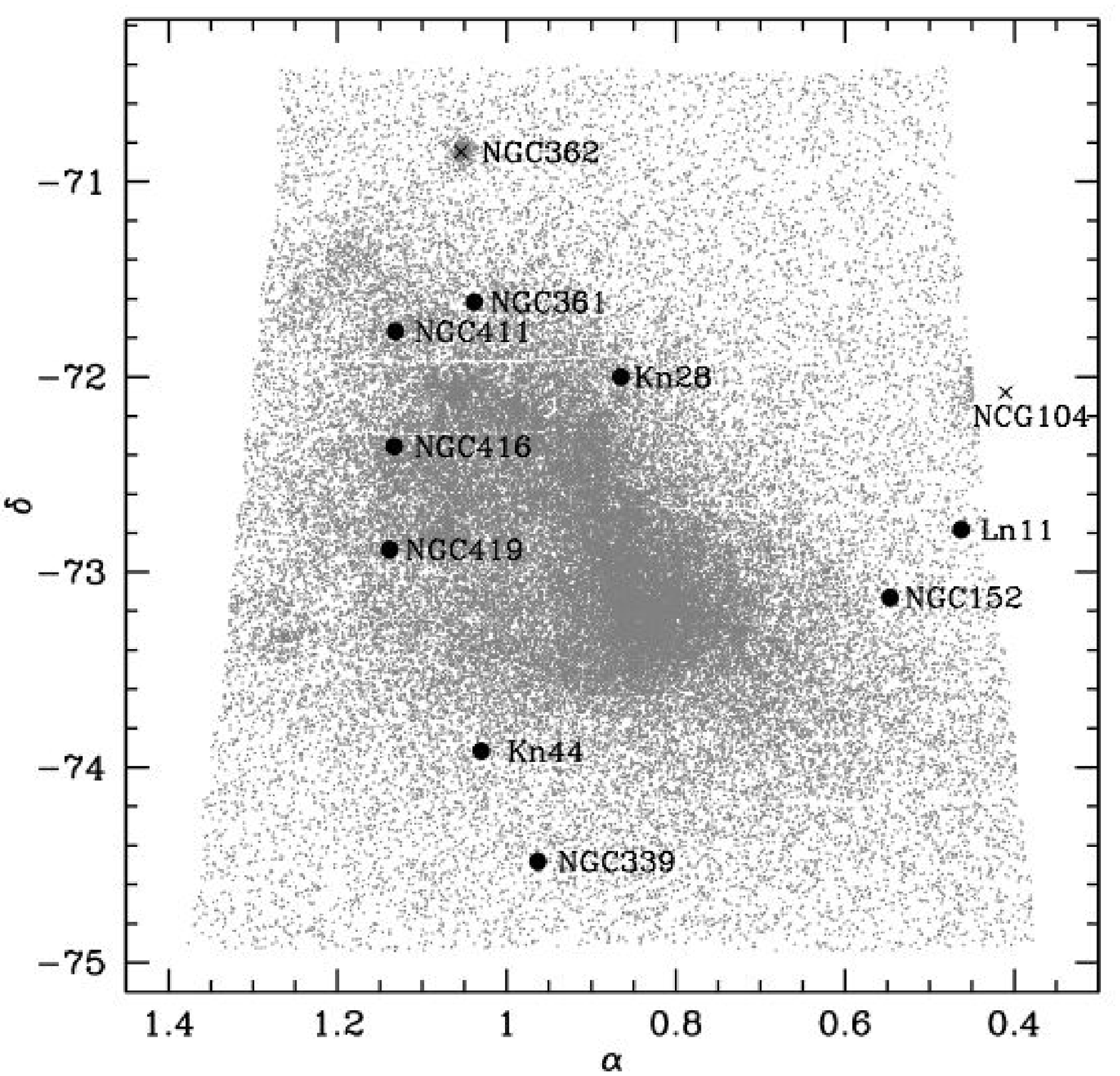}
\end{tabular}
\caption{(left) The age-metallicity relation of the SMC composed of star clusters with spectroscopic (fill circles) and photometric (open circles) metallicities. The names of the clusters of our sample are underlined (figure adopted from \cite{DC:02}). (right) Position of the clusters in the SMC with respect to the photometry by \cite{Z:02}}
\label{fig:1}       
\end{figure}

In order to eliminate as many uncertainties as possible, we obtained homogeneous spectroscopy (VLT) and deep photometry (HST/ACS) for a large sample of SMC clusters.
In this proceedings we will present the first results of the spectroscopical part of this project.
Our spectroscopical sample comprises 12 SMC clusters, 2 of which are in common with the sample of \cite{DC/H:98} (see Fig~\ref{fig:1}). Additionally, we observed 3 Galactic globular clusters for the abundance calibration.
The observations were carried out in October 2005 at the VLT at ESO/Paranal using the FORS2 spectrograph and the multi-object facility MXU. With the 1028z+29 (R$\approx 3400$) grism this configuration yields a spectral coverage of $\approx 1700 \unit{\AA}$ and a dispersion of  $\approx 0.85 \unit{\AA\, pixel^{-1}}$ in the region of the near-infrared Ca\, {\sc ii} Triplet  (CaT)
The $\lambda \lambda$8498, 8542, 8662 \AA\ lines were observed for $30 - 50 $ red giants (RGs) in each of the 15 clusters.

\subsection{Ca Triplet method}
The use of the CaT index (\SW ) in terms of magnitude difference from the horizontal branch (HB) is one of the most widely applied techniques for the derivation of abundances in individual RGs \cite{A/DC:91,C:04}. 
The CaT index is formed by the linear combination of the pseudo-equivalent width of the 3 individual lines. 
As this index is strongly dependent on stellar \Teff\ and \logg , one corrects for these effects by using the linear relation between the magnitude difference from the HB, \VVhb\ and \SW\ for stars of the same cluster (i.e. metallicity).
The resulting iso-metallicity lines in the \VVhb -\SW\ -plane are then used to derive absolute [Fe/H] abundances based on a reference [Fe/H]-scale \cite{C/G:97}.
The typical accuracy of this method is of the order of $0.2\unit{dex}$.  
\cite{C:04} have shown that the CaT method is valid over an [Fe/H] range from $-2.2\unit{dex}$ to $-0.2\unit{dex}$ and is essentially age-independent for ages ranging from 13 to 2.5 Gyr.
As our HST photometry is not fully available yet, we preliminarily use the photometric catalogue by \cite{Z:02} to estimate the $(V_{\rm HB}- V)$ values. This catalogue provides homogeneous stellar UBVI photometry of the central $18\unit{deg^{2}}$ of the SMC. 
Unfortunately, some of our target clusters lie outside of this area.

\begin{figure}[t!]
\centering
\includegraphics[height=5.1cm]{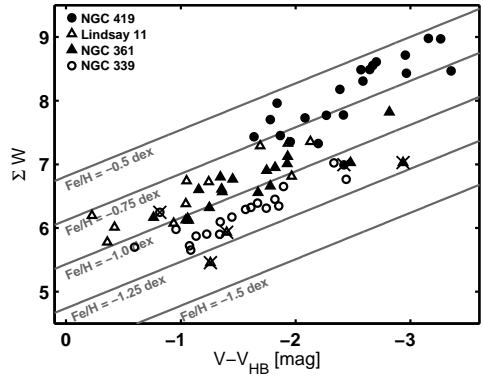}
\caption{CaT index vs. \VVhb\ for individual RGs in four of our SMC clusters. Stars of different clusters are marked by different symbols, probable non-cluster-members are crossed out.
Calibration lines in this plot have been taken from \cite{C:04}.}
\label{fig:2}       
\end{figure}

\section{Results}
From their radial velocities and position in the CMD, we selected the candidate member RGs for each cluster.
For all clusters we plotted the \VVhb -\SW\ -plane.
The majority of the datapoints for the SMC clusters follow the slope of the iso-metallicity lines.
As SCs can be presumed to be single metallicity objects without any substantial abundance spread in Fe and Ca, we assume that the very few outliers in this diagram are most likely non-cluster-members. They were ignored in the further analysis.
As an example we present the results for 4 clusters of our sample in Fig.~\ref{fig:2}.

\begin{figure}[t!]
\hskip-0.2cm\includegraphics[height=4.9cm]{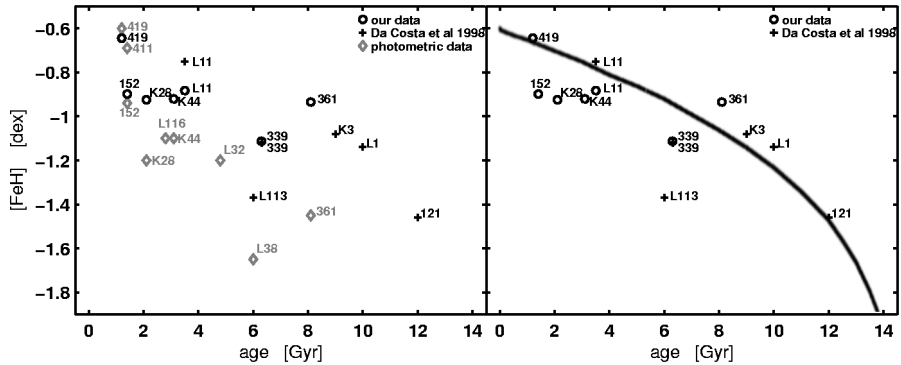}
\caption{(left) The age-metallicity data of our study are compare with those of the former study by \cite{DC/H:98} and photometric results. (right) Comparison of the spectroscopic data with the closed box model of continuous star formation (solid line) by \cite{DC/H:98}.}
\label{fig:3}       
\end{figure}

Using the [Fe/H] calibration by \cite{C:04} and adopted ages from different sources, we can highly update the age-metallicity relation by \cite{DC/H:98}.
In Fig.~\ref{fig:3} (left) we compare our results with the spectroscopic abundances  by \cite{DC/H:98} and recent photometric results \cite{FP/B:98,M:98,P:01}. 
We see excellent agreement with the results of \cite{DC/H:98} for NGC339. 
The results for L11 differ by $\approx 0.13\unit{dex}$, which is within the error range given by \cite{DC/H:98}. 
The comparison with the photometric metallicities gives a good agreement only for the 2 youngest clusters in our sample.

We compare the age-metallicity data of our study and of \cite{DC/H:98} with a simple closed box model of star formation.
Fig.~\ref{fig:3} shows that the general trend is reproduced well by this model.
Nevertheless, the flat plateau in age seen in our data between $\sim 2$ and $ 4 \unit{Gyr}$, combined  with the steep rise towards the younger end suggests that the star formation history of the SMC is probably more complicated. The inflow of unenriched gas may have played an important role.
We are awaiting the results for NGC411, L116 and L32, which lie in this region of the diagram, to see whether this pattern in the age-metallicity relation is confirmed.

The next step will be the measurement of [Fe/H] based on CaT \SW\ determinations for the remaining 5 clusters in our sample.
In total, we will have reliable spectroscopic metallicities for 10 new SMC clusters.
Furthermore, HST photometry will be available very shortly. This will enormously diminish the errors due to uncertainties and inhomogeneities in the age determination.
With these results we will obtain a well-sampled, well-defined age-metallicity relation and quantify the abundance dispersion at a given age.
The comparison of the derived age-metallicity relation with different theoretical models (closed box, leaky box, infall) will greatly improve our knowledge about the star formation history and chemical evolution of the SMC.

\bigskip
\noindent{\bf Acknowledgements}\\
A.K., E.K.G. and A.K. were supported by the Swiss National Science Foundation through the grant 200020-105260.
J.S.G. acknowledges partial research support from NSF AST-9803018 to the University of Wisconsin.

%%%%%%%%%%%%%%%%%%%%%%%% referenc.tex %%%%%%%%%%%%%%%%%%%%%%%%%%%%%%
% sample references
% "physics"
%
% Use this file as a template for your own input.
%
%%%%%%%%%%%%%%%%%%%%%%%% Springer-Verlag %%%%%%%%%%%%%%%%%%%%%%%%%%

%
% BibTeX users please use
% \bibliographystyle{}
% \bibliography{}

\printindex
\end{document}